\documentclass[sn-mathphys, Numbered]{sn-jnl}% Math and Physical Sciences Reference Style

\usepackage{graphicx}%
\usepackage{multirow}%
\usepackage{amsmath,amssymb,amsfonts}%
\usepackage{amsthm}%
\usepackage{mathrsfs}%
\usepackage[title]{appendix}%
\usepackage{xcolor}%
\usepackage{textcomp}%
\usepackage{manyfoot}%
\usepackage{booktabs}%
\usepackage{algorithm}%
\usepackage{algorithmicx}%
\usepackage{algpseudocode}%
\usepackage{listings}%
\newcommand{\apj}{Astrophys. J.}                                         		% Journal abbreviations
\newcommand{\apjs}{ApJS}
\newcommand{\apjl}{ApJ}
\newcommand{\aap}{Astron. Astrophys.}
\newcommand{\aaps}{A{\&}AS}
\newcommand{\mnras}{Mon. Not. R. Astron. Soc.}
\newcommand{\aj}{Astron. J.}
\newcommand{\araa}{ARAA}
\newcommand{\pasp}{PASP}

\newcommand{\nat}{Nature}

\theoremstyle{thmstyleone}%
\theoremstyle{thmstyletwo}%

\theoremstyle{thmstylethree}%

\begin{document}

\title[Article Title]{A Milky Way-like barred spiral galaxy at a redshift of 3}

\author*[1]{\fnm{Luca} \sur{Costantin}}\email{lcostantin@cab.inta-csic.es}
\author[1]{\fnm{Pablo G.} \sur{P\'erez-Gonz\'alez}}
\author[2]{\fnm{Yuchen} \sur{Guo}}
\author[3,4]{\fnm{Chiara} \sur{Buttitta}}
\author[2]{\fnm{Shardha} \sur{Jogee}}
\author[2]{\fnm{Micaela B.} \sur{Bagley}}
\author[5]{\fnm{Guillermo} \sur{Barro}}
\author[6]{\fnm{Jeyhan S.} \sur{Kartaltepe}}
\author[7]{\fnm{Anton M.} \sur{Koekemoer}}
\author[8,9]{\fnm{Cristina} \sur{Cabello}}
\author[4,10]{\fnm{Enrico Maria} \sur{Corsini}}
\author[11,12]{\fnm{Jairo} \sur{ M\'endez-Abreu}}
\author[13]{\fnm{Alexander} \sur{de la Vega}}
\author[14]{\fnm{Kartheik G.} \sur{Iyer}}
\author[4,10]{\fnm{Laura} \sur{Bisigello}}
\author[15]{\fnm{Yingjie} \sur{Cheng}}
\author[16]{\fnm{Lorenzo} \sur{Morelli}}
\author[17]{\fnm{Pablo} \sur{Arrabal Haro}}
\author[18,19]{\fnm{Fernando} \sur{Buitrago}}
\author[20]{\fnm{M. C.} \sur{Cooper}}
\author[21]{\fnm{Avishai} \sur{Dekel}}
\author[17]{\fnm{Mark} \sur{Dickinson}}
\author[2]{\fnm{Steven L.} \sur{Finkelstein}}
\author[15]{\fnm{Mauro} \sur{Giavalisco}}
\author[22]{\fnm{Benne W.} \sur{Holwerda}}
\author[11,12,23,24]{\fnm{Marc} \sur{Huertas-Company}}
\author[7]{\fnm{Ray A.} \sur{Lucas}}
\author[25,26]{\fnm{Casey} \sur{Papovich}}
\author[27]{\fnm{Nor} \sur{Pirzkal}}
\author[28]{\fnm{Lise-Marie} \sur{Seill\'e}}
\author[18]{\fnm{Jes\'us} \sur{Vega-Ferrero}}
\author[29]{\fnm{Stijn} \sur{Wuyts}}
\author[30]{\fnm{L. Y. Aaron} \sur{Yung}}

\affil*[1]{\orgname{Centro de Astrobiolog\'{\i}a (CAB), CSIC-INTA}, 
\orgaddress{\street{Ctra de Ajalvir km 4}, \city{Torrej\'on de Ardoz},
 \postcode{28850}, \country{Spain}}}
 
\affil[2]{\orgdiv{Department of Astronomy} 
\orgname{The University of Texas at Austin},
\orgaddress{\city{Austin}, \state{TX}, \country{USA}}}

\affil[3]{\orgname{INAF - Astronomical Observatory of Capodimonte}, 
\orgaddress{\street{Salita Moiariello 16},
\city{Naples}, \postcode{80131}, \country{Italy}}}

\affil[4]{\orgdiv{Dipartimento di Fisica e Astronomia ``G. Galilei''}, 
\orgname{Universit\`a di Padova}, 
\orgaddress{\street{Vicolo dell'Osservatorio 3}, \city{Padova}, 
\postcode{35122}, \country{Italy}}}

\affil[5]{\orgdiv{Department of Physics}, 
\orgname{University of the Pacific}, 
\orgaddress{\city{Stockton}, \postcode{90340}, \state{CA}, \country{USA}}}

\affil[6]{\orgdiv{Laboratory for Multiwavelength Astrophysics, School of 
Physics and Astronomy}, \orgname{Rochester Institute of Technology}, 
\orgaddress{\street{84 Lomb Memorial Drive},
\city{Rochester}, \postcode{14623}, \state{NY}, \country{USA}}}

\affil[7]{\orgname{Space Telescope Science Institute}, 
\orgaddress{\street{3700 San Martin Dr.}, \city{Baltimore}, 
\postcode{21218}, \state{MD}, \country{USA}}}

\affil[8]{\orgdiv{Dept. de F\'isica de la Tierra y Astrof\'isica, Fac. CC. F\'isicas}, 
\orgname{Universidad Complutense de Madrid}, 
\orgaddress{\street{Plaza de las Ciencias 1},
\city{Madrid}, \postcode{28040}, \country{Spain}}}

\affil[9]{\orgdiv{Instituto de F\'isica de Particulas y del Cosmos (IPARCOS), Fac. CC. F\'isicas}, 
\orgname{Universidad Complutense de Madrid}, 
\orgaddress{\street{Plaza de las Ciencias 1},
\city{Madrid}, \postcode{28040}, \country{Spain}}}

\affil[10]{\orgname{INAF - Osservatorio Astronomico di Padova}, 
\orgaddress{\street{Vicolo dell'Osservatorio 2},
\city{Padova}, \postcode{35122}, \country{Italy}}}

\affil[11]{\orgdiv{Departamento de Astrof\'isica}, \orgname{Universidad de La Laguna}, 
\orgaddress{\street{Avenida Astrof\'isico Francisco S\'anchez s/n},
\city{La Laguna}, \postcode{38206}, \country{Spain}}}

\affil[12]{\orgname{Instituto de Astrof\'isica de Canarias}, 
\orgaddress{\street{Calle V\'ia L\'actea s/n},
\city{La Laguna}, \postcode{38205}, \country{Spain}}}

\affil[13]{\orgdiv{Department of Physics and Astronomy}
\orgname{University of California}, 
\orgaddress{\street{900 University Ave}, \city{Riverside}, 
\postcode{92521}, \state{CA}, \country{USA}}}

\affil[14]{\orgdiv{Dunlap Institute for Astronomy \& Astrophysics}
\orgname{University of Toronto}, 
\orgaddress{\city{Toronto}, 
\postcode{M5S 3H4}, \state{ON}, \country{Canada}}}

\affil[15]{\orgname{University of Massachusetts Amherst}, 
\orgaddress{\street{710 North Pleasant Street}, \city{Amherst}, 
\postcode{01003-9305}, \state{MA}, \country{Canada}}}

\affil[16]{\orgdiv{Instituto de Astronom\'ia y Ciencias Planetarias}
\orgname{Universidad de Atacama}, 
\orgaddress{\street{Avenida Copayapu 485}, \city{Copiap\'o}, \country{Chile}}}

\affil[17]{\orgname{NSF's National Optical-Infrared Astronomy Research Laboratory}, 
\orgaddress{\street{950 N. Cherry Ave.}, \city{Tucson}, 
\postcode{85719}, \state{AZ}, \country{USA}}}

\affil[18]{\orgdiv{Departamento de F\'{i}sica Te\'{o}rica, At\'{o}mica y \'{O}ptica}, 
\orgname{Universidad de Valladolid}, 
\city{Valladolid}, \postcode{47011}, \country{Spain}}

\affil[19]{\orgdiv{Instituto de Astrof\'{i}sica e Ci\^{e}ncias do Espa\c{c}o}, 
\orgname{Universidade de Lisboa}, 
\city{Tapada da Ajuda}, \postcode{PT1349-018 }, \country{Portugal}}

\affil[20]{\orgdiv{Department of Physics \& Astronomy}
\orgname{University of California}, 
\orgaddress{\street{4129 Reines Hall}, \city{Irvine}, 
\postcode{92697}, \state{CA}, \country{USA}}}

\affil[21]{\orgdiv{Racah Institute of Physics}
\orgname{The Hebrew University of Jerusalem}, 
\orgaddress{\city{Jerusalem}, \postcode{91904}, \country{Israel}}}

\affil[22]{\orgdiv{Physics \& Astronomy Department}
\orgname{University of Louisville}, 
\orgaddress{\city{Louisville}, 
\postcode{40292}, \state{KY}, \country{USA}}}

\affil[23]{\orgdiv{Universit\'e Paris-Cit\'e, LERMA - Observatoire de Paris, PSL},
\orgaddress{\city{Paris}, \country{France}}}

\affil[24]{\orgdiv{Center for Computational Astrophysics},
\orgname{Flatiron Institute},
\orgaddress{\city{New York}, \postcode{10010}, \state{NY}, \country{USA}}}

\affil[25]{\orgdiv{Department of Physics and Astronomy}
\orgname{Texas A\&M University}, 
\orgaddress{\city{College Station}, 
\postcode{77843-4242}, \state{TX}, \country{USA}}}

\affil[26]{\orgdiv{George P.\ and Cynthia Woods Mitchell Institute for Fundamental Physics and Astronomy}
\orgname{Texas A\&M University}, 
\orgaddress{\city{College Station}, 
\postcode{77843-4242}, \state{TX}, \country{USA}}}

\affil[27]{\orgdiv{ESA/AURA Space Telescope Science Institute}}

\affil[28]{\orgdiv{Aix Marseille Univ, CNRS, CNES, LAM},
\orgaddress{\city{Marseille}, \country{France}}}

\affil[29]{\orgdiv{Department of Physics}
\orgname{University of Bath}, 
\orgaddress{\street{Claverton Down}, \city{Bath},
\postcode{BA2 7AY}, \country{UK}}}

\affil[30]{\orgdiv{Astrophysics Science Division}
\orgname{NASA Goddard Space Flight Center}, 
\orgaddress{\street{8800 Greenbelt Rd}, \city{Greenbelt}, 
\postcode{20771}, \state{MD}, \country{USA}}}

\maketitle

\textbf{The majority of massive disk galaxies in the local Universe show a stellar barred
structure in their central regions, including our Milky Way \citep{Erwin2018, PerezVillegas2020}
Bars are supposed to develop in dynamically cold stellar disks at low redshift, as the strong gas turbulence
typical of disk galaxies at high redshift suppresses or delays bar formation \citep{Tacconi2020, Zhou2020}. 
Moreover, simulations predict bars to be almost absent beyond $z = 1.5$ in the progenitors of Milky
Way-like galaxies \citep{Zana2022, Reddish2022}. Here we report observations of ceers-2112, a barred spiral galaxy
at redshift $z_{\rm phot} \simeq 3$, which was already mature when the Universe was only 2 Gyr old.
The stellar mass ($M_{\star} = 3.9 \times 10^9~M_{\odot}$) and barred morphology mean that ceers-2112 can
be considered a progenitor of the Milky Way \citep{Papovich2015, SotilloRamos2022, Xiang2022}, 
in terms of both structure and mass-assembly history in the first 2~Gyr of the Universe, and was the closest 
in mass in the first 4~Gyr. We infer that baryons in galaxies could have already dominated over dark
matter at $z \simeq 3$, that high-redshift bars could form in approximately 400~Myr and that
dynamically cold stellar disks could have been in place by redshift $z = 4–5$ (more than
12~Gyrs ago) \citep{Rizzo2020, Lelli2021}.}
\\
\\

The barred nature of ceers-2112 (right ascension = 214.97993 degrees;
declination = 52.991946 degrees; J2000.0) is identified through the
multiwavelength analysis of the James Webb Space Telescope Near Infrared
Camera (JWST/NIRCam) images taken during the first epoch (21–22
June 2022) of the Cosmic Evolution Early Release Science \citep[CEERS;][]{Finkelstein2017}
campaign. The galaxy was not classified as barred during a visual
inspection of the CEERS sample \citep{Kartaltepe2023}, owing to its low surface brightness
in the outer regions, especially at short wavelengths where the stellar
disk is barely detected. But, at longer wavelengths, ceers-2112 resembles
a spiral disk galaxy and the bar component is clearly detected by
analysing the composite image obtained by stacking all seven 
point-spread-function-convolved (PSF-convolved) NIRCam images (Fig.~1a).

The first piece of evidence for the presence of a stellar bar in ceers-2112 
is provided by the strong residuals obtained from modelling
the galaxy with a S\'ersic component (Fig.~1b). Our findings highlight
prominent features in correspondence of the spiral arms and edges
of the bar, which reveals that one morphological component is not
enough to account for the complex structure of the galaxy (for example,
twist of isophotes at small galactocentric distances and strong
residuals). Thus, we performed a multicomponent two-dimensional
(2D) photometric decomposition of ceers-2112, assuming that its
surface-brightness distribution is the sum of a double-exponential
disk and a Ferrers bar (Fig.~1c) and found that the galaxy has a stellar
bar with length $r_{\rm Ferrers} = 0.42 \pm 0.03$~arcsec (3.3 kpc). The decomposition
of the azimuthal luminosity surface-density distribution into the
Fourier $m$-components using the composite ceers-2112 image provided
the third piece of evidence that the galaxy has a prominent bar (maximum
relative amplitude of the $m = 2$ to $m = 0$ component $I_2/I_0 > 0.4$)
with strength $S_{\rm bar} = 0.23 \pm 0.01$ \citep{Aguerri2000, Athanassoula2002}. 
The $m = 2$ peak (Fig. 1d) uniquely describes the barred elongated structure and 
allowed us to rule out the possibility that the bar could be misinterpreted as spiral
arms departing from a compact bulge \citep{RosasGuevara2022}.

\begin{figure}[t]%
\centering
\includegraphics[width=\textwidth]{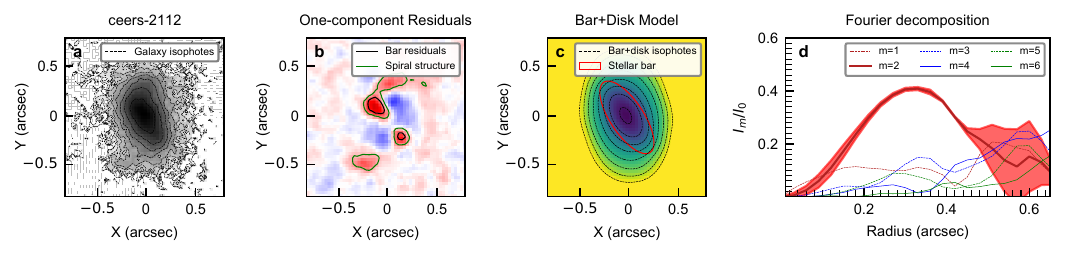}
\caption{\textbf{Morphological modeling of ceers-2112.}
\textbf{a}, Combined stack image of
ceers-2112, with isophotal contours showing an elongated barred structure in
the inner region and spiral arms departing from it. \textbf{b}, One-component S\'ersic
residuals, which highlight the bar and spiral structures (black and green
contours, respectively). \textbf{c}, Two-dimensional bar + disk model, which shows a
stellar bar of length $r_{\rm Ferrers} = 0.42 \pm 0.03$~arcsec (3.3 kpc). The bar component is
shown as a red solid line and the bar + disk isophotes are shown as black dashed
contours. \textbf{d}, Radial profiles of the relative amplitude of the odd (dashed lines)
and even (solid lines) Fourier components, derived from the deprojected
stack image of ceers-2112. The $m = 2$ mode shows a prominent bar (maximum
$I_2/I_0 > 0.4$) with strength $S_{\rm bar} = 0.23 \pm 0.01$. Shaded region represents 
$1\sigma$ confidence interval for the $m = 2$ mode.}
\end{figure}

Combining the Hubble Space Telescope Advanced Camera for Surveys
(HST/ACS), Hubble Space Telescope Wide Field Camera 3 (HST/
WFC3) and JWST/NIRCam datasets, we carefully measured ceers-2112
photometry and derived that the galaxy has a photometric redshift of
$z_{\rm phot} = 3.03_{-0.05}^{+0.04}$. Taking advantage of the unprecedented spatial resolution,
wide wavelength coverage and depth provided by JWST observations,
combined with HST datasets, we also derived the 2D spectral
energy distribution (SED) of ceers-2112 \citep{PerezGonzalez2023a}. We inferred the galaxy
star formation history (SFH) from detailed SED fitting (Fig.~2a) and
found that it has a total stellar mass of $M_{\star} = 3.9 \times 10^9 M_{\odot}$ and a 
mass-weighted age of $620_{-160}^{+150}$~Myr (Fig.~2b). 
By comparing ceers-2112 with the assembly history of Milky Way progenitors 
(Fig.~3), we demonstrated that it can be considered the furthest progenitor of the Milky
Way both in terms of structure and assembly history \citep{Xiang2022, Lucchini2023}. 
This analysis suggests that the stellar disk of ceers-2112 assembled at $z \simeq 5$ and that
the bar component formed 200~Myr later, assembling in about 400~Myr,
which provides an observational hint on the formation timescale of
bars and spiral structures at these early times. The stellar density map
built from the spatially resolved stellar population analysis (Fig.~2c)
provides an additional independent confirmation of the presence of
a stellar bar component, which has $\log(\Sigma) \simeq 8.4~M_{\odot}$~kpc$^{-2}$.

\begin{figure}[t]%
\centering
\includegraphics[width=\textwidth]{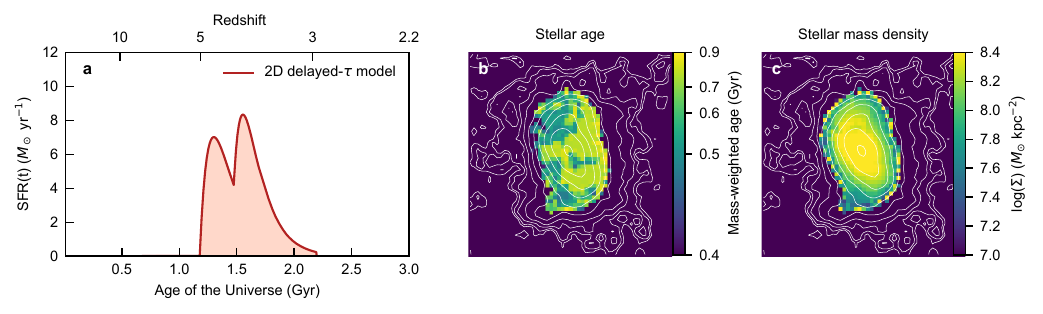}
\caption{\textbf{Stellar population properties of ceers-2112.}
\textbf{a}, Fiducial spatially-resolved SFH derived with synthesizer (delayed-$\tau$ model). 
\textbf{b}, Two-dimensional mass-weighted age map of ceers-2112. 
\textbf{c}, Stellar mass-density map of ceers-2112.
The isophotal contours of the stack image are superposed on the mass-weighted
age and mass-density maps. Maps in \textbf{b} and \textbf{c} are $53 \times 53$~px$^2$, 
which corresponds to $1.59 \times 1.59$~arcsec$^2$ ($12.5 \times 12.5$~kpc$^2$ at $z = 3.03$).}
\end{figure}

The observational discovery of barred galaxies at $z > 2$ \citep{Guo2023}, such
as ceers-2112, has strong implications for our understanding of galaxy
evolution, in particular, in the first gigayears after the Big Bang. On
the one hand, it implies that dynamically cold stellar disks could have
formed when the Universe was only a few gigayears old; on the other
hand, it puts strong constraints on the dark matter distribution in these
galaxies (with baryons dominating over dark matter).

Lambda cold dark matter ($\Lambda$CDM) models predict that galaxies at
$z > 5$ experienced a phase of gas accretion, forming stars at a very high
pace and sustaining the growth of black holes \citep{White1978, Dekel2006}. 
The baryonic cycle of this turbulent phase is balanced by strong outflows due to feedback
from active galactic nuclei and supernovae \citep{Hopkins2009, Hayward2017}. 
In the state-of-the-art cosmological simulations, different feedback implementations are
able to efficiently disperse baryons over large radial scales. However,
to build up cold stellar disks and barred galaxies at $z \gtrsim 3$, and Milky Way
systems such as ceers-2112 (Fig.~3), models should be able to reproduce
baryon-dominated disks with $M_{\star} < 10^{10}~M_{\odot}$ and net rotation at early
times. Recently, it has been shown that some massive disk galaxies
($M_{\star} > 10^{10}~M_{\odot}$) in the TNG50 cosmological simulation could have been
present as early as $z \simeq 4$ and that bars could have already started forming
at those times \citep{RosasGuevara2022}. However, despite these findings, cosmological
simulations still struggle to produce barred galaxies beyond $z > 1.5$,
especially at lower masses \citep{Zana2022, Reddish2022, Kraljic2012}.

Owing to their low entropy, galaxy disks with highly ordered rotation
are very sensitive to perturbations. However, high-z galaxies are more
gas-rich (and turbulent) than local galaxies 
\citep{Tacconi2020, Weiner2006, Genzel2006, Ubler2019} 
and gas-rich stellar disks stay near-axisymmetric much longer than 
gas-poor ones, which prevents or delays the formation of the 
bar component \citep{Athanassoula2013}. Because ceers-2112 has a mass-weighted age of 
approximately 600~Myr, the high gas fraction usually observed in 
high-z galaxies could have been rapidly consumed during the stellar disk growth 
before the stellar bar component could start developing \citep{BlandHawthorn2023}.
Thus, this result highlights the need to investigate the interplay between gas abundance and star
formation efficiency in disk galaxies at $z > 2$, which will be fundamental
in constraining the formation timescale of bars and the early evolution
of disk galaxies. Our findings allow us to speculate that ceers-2112 went
through a fast episode of gas consumption when the Universe was only
approximately 2~Gyr old ($z \simeq 4$; Fig.~2) that allowed the stellar disk to
become dynamically cold and unstable enough to allow a bar to form
and grow in less than 400~Myr \citep{BlandHawthorn2023}, which indicates a quick formation
of dynamically relaxed systems and their possible notable role in
stellar migration to the nuclear region. Using Atacama Large Millimeter/
submillimeter Array (ALMA) observations, previous works reported
the existence of cold gaseous disks at $z \simeq 5$ \citep{Rizzo2020, Lelli2021}. 
However, we observationally confirm that also stellar disks could be dynamically
cold at these early times.

\begin{figure}[t]%
\centering
\includegraphics[width=0.7\textwidth]{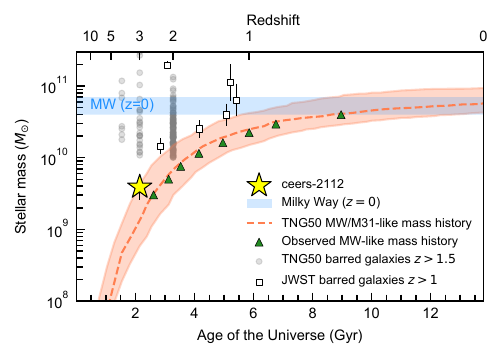}
\caption{\textbf{Mass assembly history of ceers-2112.} 
Mass-assembly history of Milky
Way (MW) and M31 analogues (dashed red line), compared with ceers-2112
(yellow star). The observed stellar masses of Milky Way progenitors are shown
as green triangles \citep{Papovich2015}. The red shaded region shows 
the $25^{\rm th}-75^{\rm th}$ percentile range \citep{SotilloRamos2022}. 
The blue shaded region stands for the observational estimates for the
stellar mass at $z = 0$ of the Milky Way \citep{Flynn2006, Licquia2015, BlandHawthorn2016}. 
Barred galaxies at $z > 1$ \citep{Guo2023, Stefanon2017}
are shown as empty squared symbols and TNG50 predictions of barred galaxies
at $z > 1.5$ \citep{RosasGuevara2022} are shown as grey dots. Error bars show 
the systematic uncertainties related to the assumptions of the SFH modelling.}
\end{figure}

\clearpage

\section*{Methods}\label{sec:methods}

\subsection*{Cosmological model}

We assume a flat $\Lambda$CDM cosmology with Hubble constant $H_0 = 67.7$~km~s$^{-1}$~Mpc$^{-1}$
and matter density $\Omega_{\rm m} = 0.310$ \citep{Planck2018}. All magnitudes are in the absolute
bolometric system.

\subsection*{Data}\label{sec:data}

JWST/NIRCam data used in this work were taken during the first epoch
(21-22 June 2022) of the CEERS program, one of 13 early release science
surveys approved for JWST Cycle 1. In particular, we focus on data from
CEERS pointing labelled NIRCam1, which is covered with seven filters:
F115W, F150W, F200W, F277W, F356W, F410M and F444W \citep{Bagley2023}. The
final mosaics in all of the filters have a pixel scale of 0.03~arcsec~px$^{-1}$
and a PSF full-width at half-maximum (FWHM) ranging from 0.066
to 0.161~arcsec, reaching a point-source limiting magnitude ($5\sigma$) of
approximately 29~mag \citep{Bagley2023, Finkelstein2023}. The root mean square astrometric
alignment quality is less than approximately $5-10$~mas per source
between NIRCam filters \citep{Bagley2023}. In Extended Data Fig.~1, 
we show the cutouts of ceers-2112 in all NIRCam bands.
For studying the morphology of ceers-2112, we built a stack image
combining all seven NIRCam bands. We converted individual images
in counts and PSF convolved them to match the angular resolution of
the F444W image. Empirical PSFs for the CEERS datasets are created as
described in \citep{Finkelstein2023}, whereas kernels to match bluer PSFs to F444W are
created using the pypher Python-based routine \citep{Boucaud2016}. Finally, we combined
all PSF-convolved images using the ccdproc.combine v.2.4.0 astropy
image reduction package \citep{Craig2023}.
For studying the stellar population properties of ceers-2112, we
extended the NIRCam wavelength baseline with HST images (F606W,
F814W, F125W, F140W and F160W) from the Cosmic Assembly
Near-infrared Deep Extragalactic Legacy Survey (CANDELS) collaboration 
\citep{Grogin2011, Koekemoer2011}.
These data were recalibrated by the CEERS team and drizzled
to match the same angular scale of the JWST observations \citep[v.1.9;][]{Bagley2023}.

\subsection*{Morphology}\label{sec:morphology}

We analysed the morphology of ceers-2112 by modelling its surface-brightness
distribution to characterize its structural components using
four different diagnostics: (1) isophotal analysis; (2) Fourier decomposition;
(3) one-component S\'ersic photometric modelling; and
(4) two-component bar + disk photometric modelling.
Firstly, we considered the radial surface-brightness profile of 
ceers-2112 and modelled its isophotes with the photutils.isophote astropy
package \citep{Jedrzejewski1987, Bradley2020} 
using three short wavelength bands (that is, F115W, F150W
and F200W; Extended Data Fig.~2a,d) and three long wavelength bands
(that is, F277W, F356W and F444W; Extended Data Fig.~2b,e). We created
ellipticity and position angle profiles by keeping the centre fixed
to the average value measured in the inner region of the galaxy. Then,
we checked that ceers-2112 satisfied the criteria 
\citep{Guo2023,Jogee2004, Marinova2007} of hosting a putative
bar-dominated region: (1) the galaxy became elongated in the bar
region (ellipticity $\epsilon > 0.25$) and the position angle remained almost
constant along the bar ($| \Delta$ position angle$|$ $< 15^{\circ}$); and (2) the ellipticity
dropped in the outer region of the galaxy ($\Delta \epsilon = 0.1$), where the disk component
dominates. Our findings suggest the presence of a bar, which
appears more prominent at longer wavelengths, with an ellipticity
always greater than 0.4 up to a radius of $r \simeq 0.45$~arcsec. It is worth noting
that the analysis of individual bands is complicated by the presence of
spiral arms, which could drive the mild change in position angle (and
slightly affect the ellipticity) in the outskirts of the galaxy. As a caveat,
the disk component is very mildly detected (in particular, in short wavelength
bands), leading to a small change in bar-to-disk $\epsilon$ and position
angle. For this reason, we decided to further analyse the morphology
of ceers-2112 using the combined image obtained by stacking all
NIRCam filters, to increase the final signal-to-noise ratio, in particular,
in the outskirts of the galaxy. In the combined image (Extended Data
Fig.~2c,f), our analysis showed an inner bar-dominated region ($\epsilon > 0.4$,
$\Delta$ position angle $< 15^{\circ}$), a region where mild spiral arms develop from
the barred structure and the outer disk-dominated region, where the
ellipticity and position angle drop \citep{Wozniak1995, MunozMateos2013}.

Secondly, we analysed the deprojected combined stack image of
ceers-2112 and decomposed its azimuthal luminosity surface-density
distribution into the Fourier $m$-components \citep{Buttitta2022}. To project the galaxy into
the face-on view keeping the flux preserved, the image was stretched
along the disk minor axis by a factor of cos($i_{\rm disk}$)$^{-1}$, where $i_{\rm disk}$ is the disk
inclination derived from the disk ellipticity. In particular, from the isophotal
fitting of the combined image we derived $\epsilon_{\rm disk} = 0.23$ ($i_{\rm disk} = 41^{\circ}$),
taking the median values in the outer isophotes ($0.6 < r < 0.7$~arcsec)
where the influence of the bar is negligible (Extended Data Fig.~2c,f).
In Fig.~1d, we show the radial profiles of the relative amplitude of the
$m = (2, 4, 6)$ components. In particular, the $m = 2$ component shows
the characteristic behaviour of bars \citep{Aguerri2000, RosasGuevara2022}: increasing with radius (with
a prominent peak $I_2/I_0 > 0.4$) and then decreasing in the disk region.
The phase angle $\phi_2$ of the $m = 2$ component is quite constant in the
bar region ($|\Delta \phi_2 | < 10^{\circ}$ with respect to the $I_2/I_0$ peak), which provides
an additional confirmation of the presence of the bar component. We
further tested our findings by repeating the Fourier decomposition
assuming both different position angles ($\Delta$ position angle $\pm 5^{\circ}$) and
inclinations ($\Delta i \pm 5^{\circ}$) for the galaxy (eight different configurations). No
systematics were found in the bar identification due to galaxy deprojection
effects. Furthermore, it is worth reporting that the bar/interbar
intensity contrast based on the Fourier decomposition provides results
about the length and strength of the bar that are consistent with those
of the $m = 2$ Fourier analysis \citep{Aguerri2000, Aguerri2015}. It is also worth noting that our Fourier
analysis allows us to rule out the possibility that the stellar bar could
be misled by spiral arms developing from a compact bulge. Indeed,
this latter case would not produce an $m = 2$ peak in the inner region
of the galaxy \citep{RosasGuevara2022}.

Thirdly, to disentangle the contribution to the surface brightness of
bar and spiral arms, we modelled the galaxy with a single S\'ersic component
and looked at the residual image (Fig.~1b). We used the Python package
statmorph \citep{RodriguezGomez2019} to retrieve both the parametric and non-parametric
morphology of the galaxy. The best-fitting model provides a quite low
S\'ersic index $n = 0.65$ (disky galaxy), with the residual image highlighting
prominent features in correspondence of the spiral arms and edges
of the bar component. Our findings suggest that the one-component
S\'ersic model is not sufficient to describe the complex morphology
of ceers-2112.

Finally, we perform a 2D photometric decomposition of ceers-2112
using the galaxy surface photometry 2D decomposition algorithm
\citep[GASP2D;][]{MendezAbreu2008, MendezAbreu2014}. 
We model the galaxy (Fig.~2c) by assuming that its
surface-brightness distribution is the sum of a double-exponential
disk \citep{vanderKruit1979} and a Ferrers bar \citep{Aguerri2009}. 
GASP2D returns the best-fitting values of the
structural parameters of each morphological component by minimizing
the $\chi^2$ after weighting the surface brightness of the image pixels
according to the variance of the total observed photon counts due
to the contribution of both galaxy and sky (Extended Data Fig.~4c,d).
Because GASP2D does not fit the spiral arm components, we mask
them to avoid possible contamination in retrieving the ellipticity and
position angle of the bar. The mask for the 2D bar + disk decomposition
is built by growing the spiral arms residuals, excluding the bar region.
Because the formal errors obtained from the $\chi^2$ minimization are usually
not representative of the real errors, we estimated the uncertainties on
the bar and disk parameters by analysing a sample of images of mock
galaxies built with Monte Carlo simulations \citep{Costantin2017}.

As a caveat, as the composite stack image covers the wavelength
range from the rest-frame ultraviolet to near infrared, dust attenuation
and spatially variable younger stellar populations may result
in a composite light distribution that does not follow the stellar distribution.
To support our analysis, we then created combined short
wavelength (F115W, F150W and F200W) and long wavelength (F277W,
F356W and F444W) stack images, following the procedure described in
the previous section. The short wavelength stack image was convolved
to F200W, whereas the long wavelength stack image was convolved
to F444W. The isophotal analysis of short wavelength stack and long
wavelength stack images is shown in Extended Data Fig.~3c,d. We see a
similar trend at short and long wavelengths, with two main differences:
(1) the position angle is almost constant in the long wavelength stack
image, although it shows a mild variation in the short wavelength stack
image, making the identification of the bar less clear in the 
ultraviolet--optical rest-frame regime; and (2) the signal-to-noise ratio, in particular,
in the inner and outer regions, is very low in the short wavelength stack
image with respect to the long wavelength stack image. The Fourier
analysis of these images is shown in Extended Data Fig.~3e. We see that
the bar component is clearly detected at longer wavelengths ($m = 2$
component stronger than any other component), whereas, at shorter
wavelengths, we see both prominent $m = 1$ and $m = 2$ components. This
is due to the non-asymmetry (lopsidedness) of the elongated structure
seen at short wavelengths. Again, while the evidence for the bar
structure is present both at short and long wavelengths, the bar is more
evident in the redder bands, as expected from near infrared studies in
the local Universe \citep{Eskridge2000}.

For the reasons described above, we based our main analysis on
the image obtained by combining all seven NIRCam bands. Our morphological
analysis provided four independent estimations of the bar
length: (1) $R_{\rm bar,1} = 0.49 \pm 0.09$~arcsec, from the outer radius of the FWHM
of $I_2/I_0$ \citep{Ohta1990}; (2) $R_{\rm bar,2} = 0.44 \pm 0.04$~arcsec, from the outer radius of
the FWHM of the bar/interbar contrast \citep{Aguerri2015}; (3) $R_{\rm bar,3} = 0.49 \pm 0.02$~arcsec,
from the radius at which there is the first minimum after the (deprojected)
ellipticity peak \citep{Wozniak1995}; and (4) $R_{\rm bar,4} = 0.42 \pm 0.03$~arcsec, from the
Ferrers bar modelling \citep{Aguerri2009}.

\subsection*{Redshift estimation}\label{sec:redshift}

We carefully measured the ACS, WFC3 and NIRCam photometry with
the rainbow code \citep{PerezGonzalez2008, Barro2011}, 
using small elliptical apertures (radius of 0.44~arcsec; 
$\epsilon = 0.35$) to retrieve reliable colours and avoid possible photometric
contamination by a foreground extended source ($z_{\rm phot} = 1.1$;
projected distance of approximately 3.5~arcsec). Then, we measured
the photometry on slightly larger apertures (radius of 0.84~arcsec)
to obtain the integrated emission. Finally, we normalized the SED
measured on small apertures using the median difference of the flux
measured in the small and large apertures. Photometric errors were
estimated by measuring the background noise locally around 
ceers-2112, which accounted for correlated noise introduced by drizzling
the ACS, WFC3 and NIRCam images \citep{PerezGonzalez2008, PerezGonzalez2023b}.

The fiducial photometric redshift was derived using EAZYpy \citep{Brammer2008} including
the tweak\_fsps\_QSF\_12\_v.3 set of 12 Flexible Stellar Population
Synthesis templates \citep{Conroy2010, Barro2023}. The combined HST + JWST SED, the values of
the photometric redshift and the corresponding probability density
functions are shown in Extended Data Fig.~5. We further tested the
photometric redshift estimation against different codes (that is, Dense
Basis \citep{Iyer2019}; Prospector \citep{Johnson2021}) 
and found consistent results (see Extended Data Fig.~5, inset panel).

\subsection*{Stellar population properties}\label{sec:stepop}

We derived the fiducial spatially resolved SFH of ceers-2112 with synthesizer 
\citep{PerezGonzalez2008}, assuming delayed-exponential SFHs. We adopt timescale
values $\tau$ between 100~Myr and 5~Gyr, ages between 1~Myr and the
age of the Universe at the redshift of ceers-2112, the entire set of discrete
metallicities provided by the \citet{Bruzual2003} models, a \citet{Calzetti2000} 
attenuation law with $V$-band extinction values between 0 and
5~mag and a \citet{Chabrier2003} initial mass function. The nebular continuum
and emission lines were added to the models \citep{PerezGonzalez2008}.
We further tested the systematics related to the stellar population
modelling by deriving the integrated stellar population properties of
ceers-2112, using both parametric and non-parametric SFHs (Extended
Data Fig.~6 and Extended Data Table~1). For this purpose, we fitted the
integrated HST + NIRCam photometry using the Fitting and Assessment
of Synthetic Templates (FAST) code \citep{Kriek2009}, 
Dense Basis \citep{Iyer2019} and Prospector \citep{Johnson2021}.
For the FAST algorithm, we assumed an exponentially declining star
formation history. We used \citet{Bruzual2003} stellar population
synthesis models, \citet{Calzetti2000} extinction law with attenuation
$0 < A_V < 4$~mag, and a \citet{Chabrier2003} initial mass function. For Dense Basis,
we use a uniform prior for the stellar mass $\log(M_{\star}/M_{\odot})$ between 7 and
12, uniform prior for the metallicity $\log(Z/Z_{\odot})$ between $-1.5$ and 0.25,
a \citet{Calzetti2000} attenuation law with exponential prior and $V$-band
extinction values between 0 and 4~mag, and a \citet{Chabrier2003} initial mass
function. For Prospector \citep{Johnson2021,Leja2019}, we used both a delayed-exponential
and a non-parametric SFH. For the $\tau$-model, we used stellar ages ranging
between 1~Myr and the age of the Universe at the redshift of ceers-
2112 and the star formation scale in the range $0.1 < \tau < 20$~Gyr. For the
non-parametric model, we used an SFH with the continuity prior \citep{Leja2019}. We
adopted five lookback time bins in this fit, with the star formation rate
being constant within each bin. The first bin was fixed at $0 < t < 30$~Myr to
capture the recent episodes of star formations. We used uniform priors
on all of the following parameters: stellar mass $\log(M_{\star}/M_{\odot})$ between 5
and 12, metallicity $\log(Z/Z_{\odot})$ between $-1.5$ and 0.5 and effective $V$-band
optical depth between 0 and 5~mag. We adopted the \citet{Chabrier2003} initial
mass function and the \citet{Calzetti2000} dust attenuation law.

\backmatter

\section*{Extended data figures and tables}

\begin{figure}[h]% 
\centering
\includegraphics[width=\textwidth]{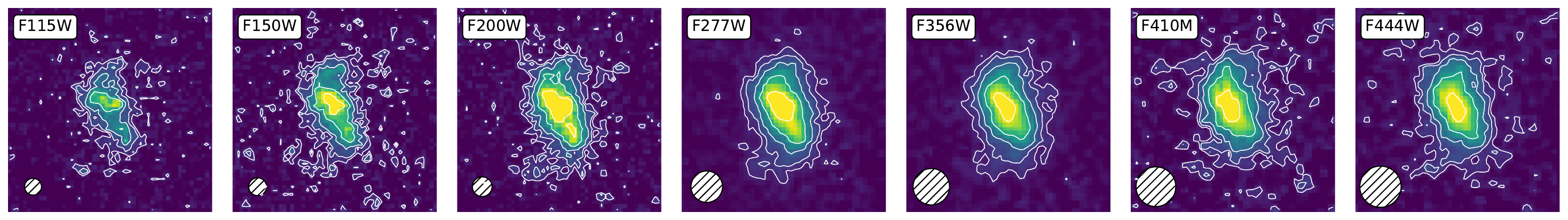}
\caption{\textbf{Multiwavelength view of ceers-2112.}
Postage stamps of ceers-2112 (RA = 214.97993 degrees; DEC = 52.991946 degrees; J2000.0) in all
NIRCam filters used in this work. The cutouts are $53 \times 53$~px$^2$, which corresponds
to $1.59 \times 1.59$~arcsec$^2$ ($12.5 \times 12.5$~kpc$^2$ at $z = 3.03$). We report the angular
resolution as 2 $\times$ FHWM of the PSF and the isophotal contours (white solid lines). 
}
\end{figure}

\begin{figure}[h]%
\centering
\includegraphics[width=\textwidth]{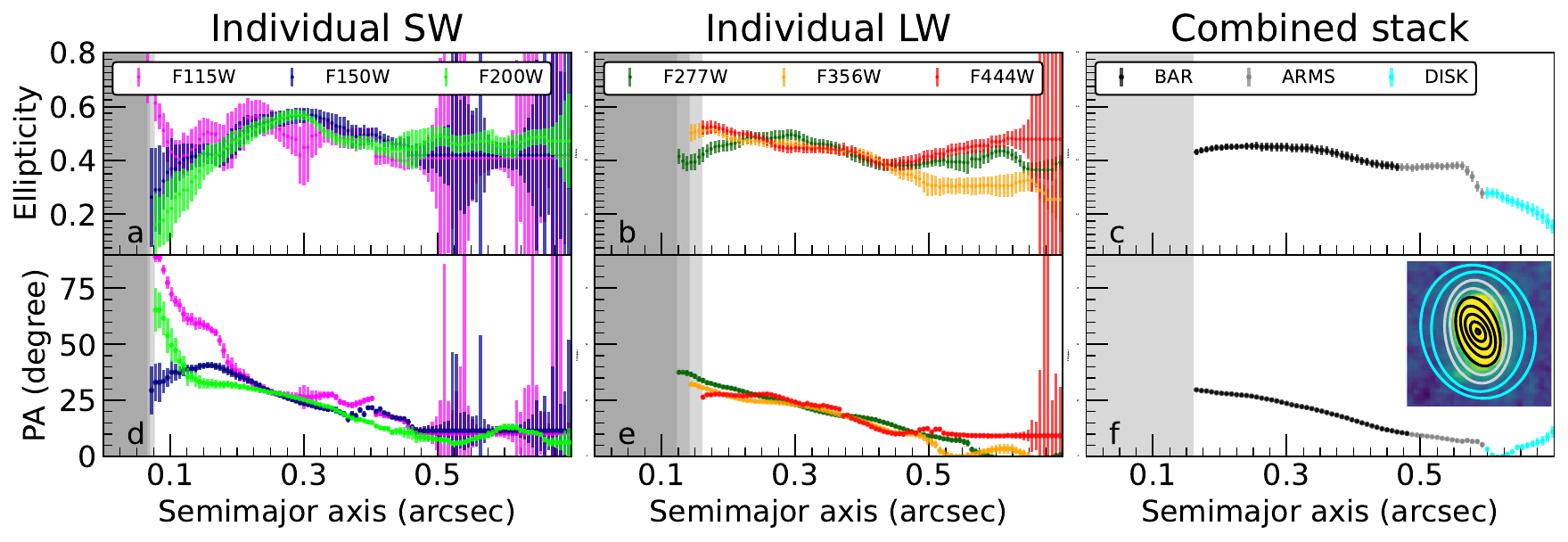}
\caption{\textbf{Isophotal analysis of ceers-2112.}
In each panel, the shaded regions mark the size of the PSF FWHM in the different bands, while
error bars show the $1\sigma$ standard deviation of each point. \textbf{a}, Radial profiles of
ellipticity derived from the isophotal analysis of ceers-2112 in the F115W (pink),
F150W (blue) and F200W band (light green). \textbf{b}, Radial profiles of ellipticity
derived from the isophotal analysis of ceers-2112 in the F277W band (dark
green), F356W (orange) and F444W band (red). \textbf{c}, Radial profiles of ellipticity
derived from the isophotal analysis of ceers-2112 in the combined stack image
(all seven NIRCam filters). The region of the bar, spiral arms and outer disk are
shown as black, grey and cyan datapoints. \textbf{d}, As panel \textbf{a}, but for the position
angles. \textbf{e}, As panel \textbf{b}, but for the position angles. \textbf{f}, As panel \textbf{c}, but for the position
angles. The inset panel shows some of the ellipses superposed to the composed
stack image ($1.59 \times 1.59$~arcsec$^2$).
}
\end{figure}

\begin{figure}[h]%
\centering
\includegraphics[width=\textwidth]{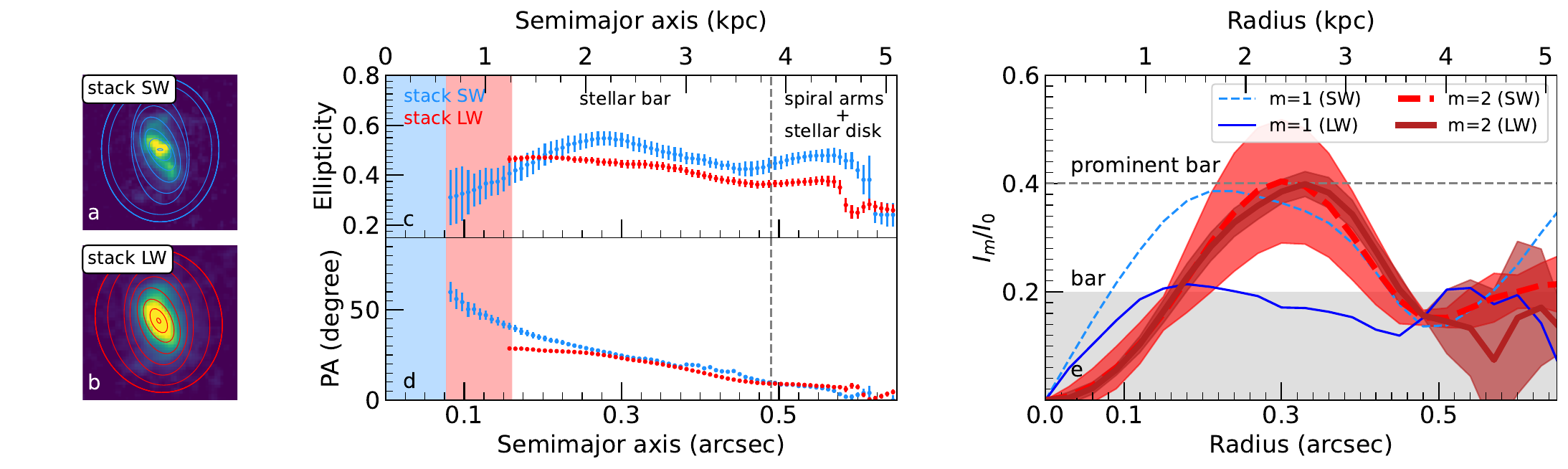}
\caption{\textbf{Isophotal and Fourier analysis of SW and LW stack images of ceers-2112.}
\textbf{a}, Postage stamp of the stack SW image (F115W, F150W
and F200W) with some of the ellipses superposed ($1.59 \times 1.59$~arcsec$^2$). 
\textbf{b}, Postage stamp of the stack LW image (F277W, F356W and F444W) with some of the
ellipses superposed ($1.59 \times 1.59$~arcsec$^2$). \textbf{c}, Radial profiles of ellipticity derived
from the isophotal analysis of ceers-2112 in the stack SW image (blue) and stack
LW image (red). The shaded regions mark the size of the PSF FWHM in the
different bands, while error bars show the $1\sigma$ standard deviation of each point.
\textbf{d}, As \textbf{c}, but for the position angles. \textbf{e}, Radial profiles of the relative amplitude
of the $m = 1$ (blue lines) and $m = 2$ (red lines; shaded regions: $1\sigma$ confidence
intervals) Fourier components derived from the deprojected combined SW
image (dashed lines) and the deprojected combined LW image (solid lines) of ceers-2112.
}
\end{figure}

\begin{figure}[h]%
\centering
\includegraphics[width=\textwidth]{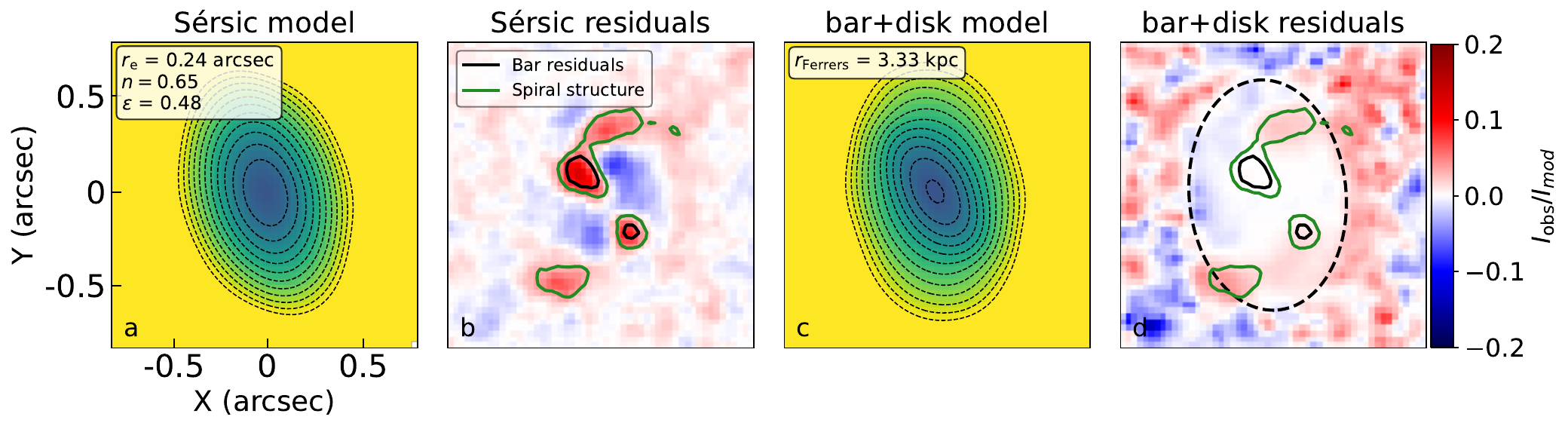}
\caption{\textbf{Parametric morphological modeling of ceers-2112.}
\textbf{a}, One-component S\'ersic model of ceers-2112. \textbf{b}, One-component S\'ersic
residuals, which highlight the bar and spiral structures (black and green
contours, respectively) \textbf{c}, Two-dimensional bar+disk model, which shows a
stellar bar of length $r_{\rm Ferrers} = 0.42 \pm 0.03$~arcsec (3.3 kpc). \textbf{d}, Two-dimensional
bar+disk model residuals. The bar and spiral structures (black and green
contours, respectively) are superposed to the image. The black dashed line
marks the break radius of the double-exponential disk model, where the
surface brightness of the model rapidly declines.
}
\end{figure}

\begin{figure}[h]%
\centering
\includegraphics[width=\textwidth]{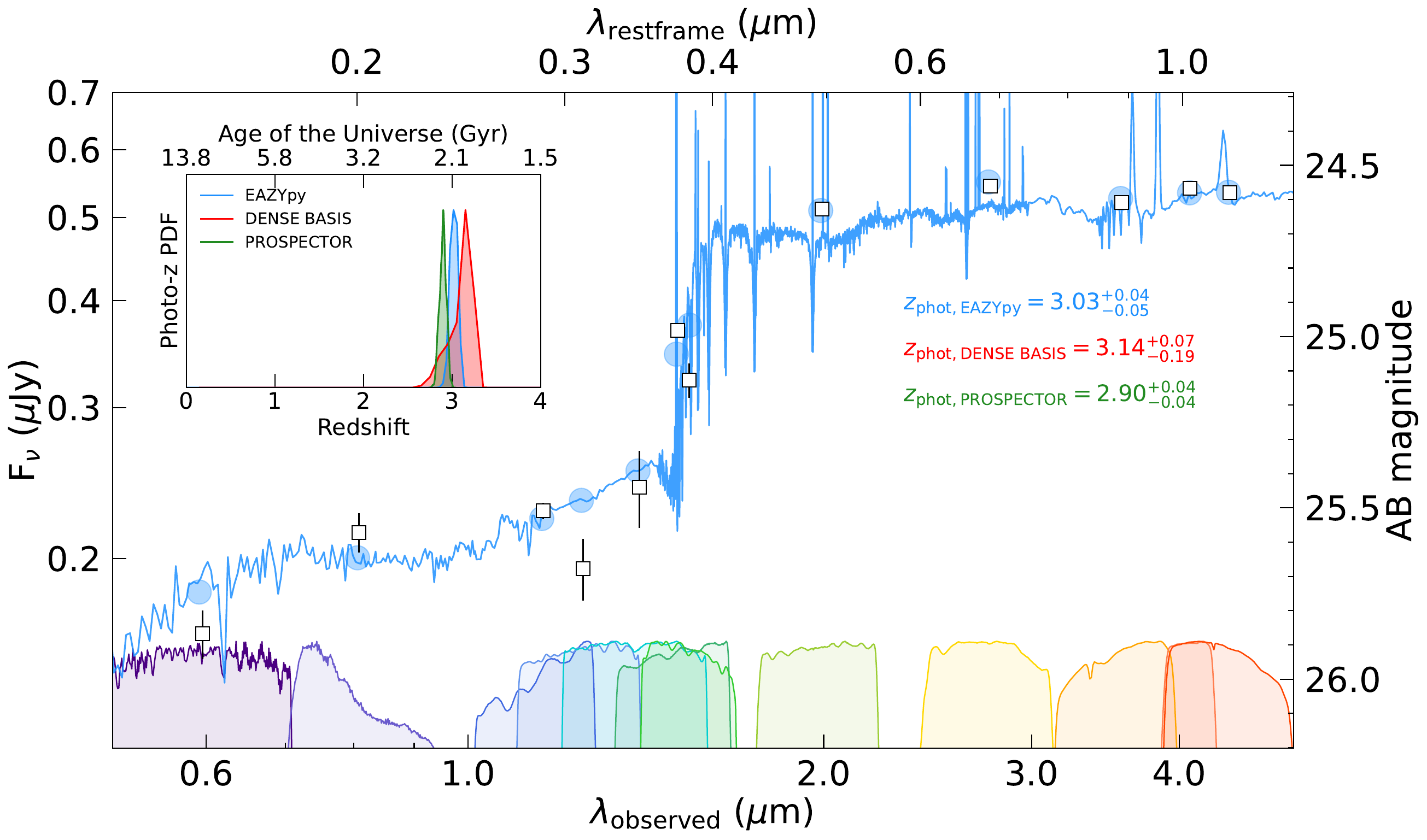}
\caption{\textbf{SED and redshift of ceers-2112.} 
Black empty squares
(blue circles) denote our fiducial (model) photometry from HST/ACS + WFC3
and JWST/NIRCam instruments, respectively. The EAZYpy model spectrum is
shown in blue. Error bars show the $1\sigma$ standard deviation of each point. The
inset plot shows the P(z) distributions.
}
\end{figure}

\begin{figure}[h]%
\centering
\includegraphics[width=0.6\textwidth]{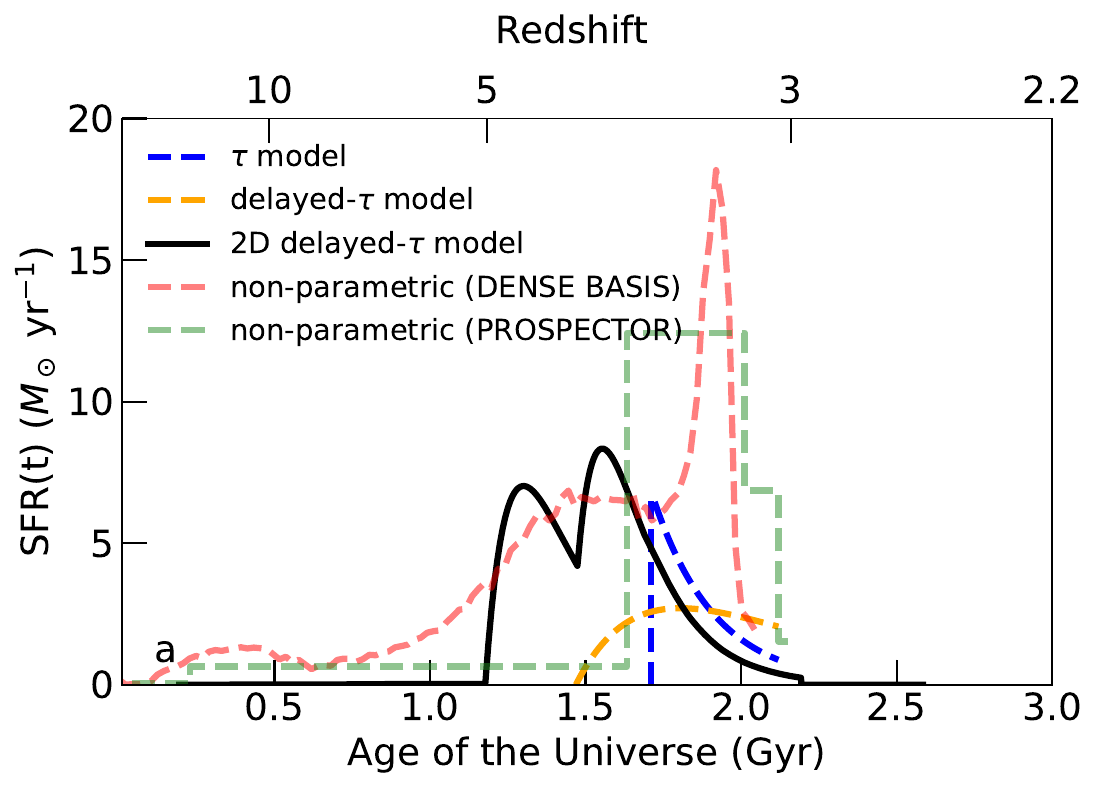}
\caption{\textbf{SFH modeling of ceers-2112.}
Comparison of different
model assumptions: exponentially-declining SFH ($\tau$-model; FAST code; blue
dashed line); delayed exponentially-declining SFH (delayed-$\tau$ model; Prospector;
gray dashed line); two-dimensional delayed exponentially-declining SFH
(2D delayed-$\tau$ model; synthesizer; red solid line); non-parametric SFH (Dense
Basis and Prospector; orange and green dashed lines, respectively).
}
\end{figure}

\begin{table}[h]
\caption{Mass and $SFR_{50}$ of ceers-2112 derived from different model assumptions.}
\begin{tabular}{cccccc}
\toprule
& 2D delayed-$\tau$ & delayed-$\tau$  & $\tau$-model & non-parametric & non-parametric\\
\midrule
code    						& \texttt{synthesizer} 	& \texttt{PROSPECTOR}  & \texttt{FAST} 		& \texttt{PROSPECTOR} 	& \texttt{DENSE~BASIS} 	\\
$M_{\star}$ ($M_{\odot}$)  		& $3.9\times10^9$  		& $4.9\times10^9$		& $2.6\times10^9$   	& $4.5\times10^9$   		& $4.6\times10^9$   		\\
$\log(SFR_{50})$ ($M_{\odot}$ yr$^{-1}$) & $-0.54$   			& $0.33$ 				& $-0.01$  		& $0.73$  				& $0.31$   			\\
\botrule
\end{tabular}
\end{table}

\clearpage

\bmhead{Acknowledgments}

We thank S.~Roca-F\`abrega and E. Borsato for their comments. L.C.
acknowledges financial support from the Comunidad de Madrid under Atracci\'on de Talento
grant no. 2018-T2/TIC-11612. L.C. and P.G.P.-G. acknowledge support from grant nos. PGC2018-
093499-B-I00, PID2022-139567NB-I00 and MDM-2017-0737 Unidad de Excelencia ‘Maria de
Maeztu’ Centro de Astrobiolog\'ia (INTA-CSIC) funded by the Spanish Ministry of Science and
Innovation/State Agency of Research MCIN/AEI/ 10.13039/501100011033, FEDER, UE. C.C.
acknowledges support from grant nos. PRE2019-087503 and PID2021-123417OB-I00 funded
by MCIN/AEI/ 10.13039/501100011033 ‘ESF Investing in your future’ and ‘ERDF A way of making
Europe’, respectively. J.M.A. acknowledges the support of the Viera y Clavijo Senior program
funded by ACIISI and ULL and the support of the Agencia Estatal de Investigaci\'on del
Ministerio de Ciencia e Innovaci\'on (MCIN/AEI/10.13039/501100011033) under grant nos.
PID2021-128131NB-I00 and CNS2022-135482 and the European Regional Development Fund
(ERDF) ‘A way of making Europe’ and the ‘NextGenerationEU/PRTR’. F.B. and J.V.F. acknowledge
the support from grant no. PID2020-116188GA-I00 by the Spanish Ministry of Science and
Innovation, and F.B. also acknowledges grant no. PID2019-107427GB-C32. Support for this
work was provided by NASA through grant no. JWST-ERS-01345 awarded by the Space
Telescope Science Institute, which is operated by the Association of Universities for Research
in Astronomy, Inc., under NASA contract NAS 5-26555.

\clearpage

%% BioMed_Central_Bib_Style_v1.01

\end{document}